# On the spheroidized carbide dissolution and elemental partitioning in a high carbon bearing steel 100Cr6

Wenwen Song[1]  Pyuck-Pa Choi[2]  Gerhard Inden[2]  Ulrich Prahl[1]  Dierk Raabe[2]  Wolfgang Bleck[1]

[1] Department of Ferrous Metallurgy, RWTH Aachen University, Germany
[2] Max-Planck-Institut für Eisenforschung GmbH, Düsseldorf, Germany

**Abstract**

We report on the characterization of high carbon bearing steel 100Cr6 using electron microscopy and atom probe tomography in combination with multi-component diffusion simulations (DICTRA). Scanning electron micrographs show that around 14 vol.% spheroidized carbides are formed during soft annealing and only 3 vol.% remain after dissolution into the austenitic matrix by austenitization at 1123 K (850 °C) for 300 s. The spheroidized particles are identified as $(Fe, Cr)_3C$ by transmission electron microscopy. Atom probe analyses reveal the redistribution and partitioning behaviors of elements, i.e. C, Si, Mn, Cr, Fe in both, the spheroidized carbides and the bainitic matrix in the sample isothermally heat-treated at 773 K (500 °C) after austenitization. A homogeneous distribution of C and gradual gradient of Cr was detected within the spheroidized carbides. Due to its limited diffusivity in $(Fe, Cr)_3C$, Cr exhibits a maximum concentration at the surface of spheroidized carbides (16 at.%) and decreases gradually from surface towards the core down to a level of about 2 at.%. The atom probe results also indicate that the partially dissolved spheroidized carbides during austenitization may serve as nucleation sites for intermediate temperature cementite within bainite, which results in a relatively softer surface and harder core in spheroidized particles. This microstructure may contribute to the good wear resistance and fatigue properties.

1. Introduction

Controlling precipitation in steels for enhancing their wear resistance and fatigue properties has been a subject of intense and long-lasting research with a variety of advanced experimental and modeling approaches [1-14]. 100Cr6 steels with the basic composition of 1 wt.% C and 1.5 wt.% Cr are among the most extensively used materials for bearings in the industry. As these materials encounter substantial thermomechanical loading in service, the material and heat treatment design should meet the requirements of high fatigue and wear resistance as well as an outstanding combination of strength and toughness.

A special soft annealing treatment, spheroidization, produces a mixed microstructure of relatively coarse spheroidized cementite particles embedded in ferrite, which facilitates machining, warm and cold forming of the material. This microstructure can be subjected to further heat treatment to achieve a final martensitic or bainitic microstructure. Spheroidization kinetics has been long known to be influenced by carbon and chromium diffusion and their concentrations. Higher carbon concentration promotes the spheroidization process, because it provides a higher number density of nucleation sites. Chromium reduces the inter-lamellar spacing of pearlite, which is often the starting structure for spheroidization process [15]. Spheroidization in 100Cr6 has a great influence on the subsequent bainitic and pearlitic transformation. By varying the spheroidization process parameters, namely the holding time and temperature, the dissolution kinetics can be controlled. In this way, the desired content of spheroidized carbides and the distribution of carbon content in both spheroidized carbides and ferrite can be achieved.

In the present work, we characterized the spheroidized carbides with respect to the morphology, crystallography, phase fraction, size distribution, transformation kinetics and chemical gradients of different elements using electron microscopy and atom probe tomography in combination with multi-component diffusion simulations (DICTRA). Atom probe tomography (APT) was employed to study the elemental distributions in spheroidized carbides and bainite [16]. The partitioning behavior of carbon and other alloy elements across the phase boundaries are discussed, with an emphasis on the effect of Cr, Mn and Si on the growth kinetics of cementite.

## 2. Experimental

The chemical composition of the studied steel 100Cr6 is given in Table 1. The steel is mainly alloyed with Cr and microalloyed with Mo. Si and Mn contents are at a low level, while the Al content is almost negligible. The N content in the steel is 75 ppm.

Table 1. Chemical composition of the investigated steel 100Cr6

| Element | C | Si | Mn | P | S | Cr | Mo | Ni | Cu | Al |
|---|---|---|---|---|---|---|---|---|---|---|
| **wt.%** | 0.967 | 0.30 | 0.23 | 0.003 | <0.001 | 1.38 | 0.02 | 0.07 | 0.05 | 0.026 |
| **at.%** | 4.325 | 0.58 | 0.23 | 0.005 | <0.002 | 1.43 | 0.01 | 0.07 | 0.04 | 0.052 |

The heat treatment cycle and investigated conditions (HTC) are illustrated in Table 2 and Fig. 1. After hot forging, the material was soft annealed industrially and cooled down to form a spheroidized microstructure (HTC1). Starting with the spheroidized microstructure, the samples were heated up at a rate of 3.3 K/s and austenitized at 1123 K (850 °C) for 300 s. After austenitization, two heat treatment routes were performed. One was quenching to room temperature in Ar (HTC2) and the other was rapidly cooling to 773 K (500 °C) at a rate of 55 K/s. At 773 K (500 °C), the samples were isothermally held for 1200 s in order to form a bainitic microstructure (HTC3) with subsequent air cooling. Austenitization and bainitization were performed in a Bähr 805A dilatometer, where the dimension of the specimen was Φ3 mm × 10 mm.

Table 2. Heat Treated Conditions(HTC) of 100Cr6 for investigation

| HTC No. | Heat Treatment Conditions (HTC) |
|---|---|
| **HTC 1** | Spheroidization + cooling to room temperature (RT) |
| **HTC 2** | Spheroidization + austenitization at 1123 K (850 °C) for 300 s + quenching to room temperature (RT) in Ar |
| **HTC 3** | Spheroidization + austenitization at 1123 K (850 °C) for 300 s + Isothermal holding at 773 K (500 °C) for 1200 s + cooling to room temperature (RT) in air |

Microstructural characterization was performed using scanning (SEM) and transmission electron microscopy (TEM). TEM studies were done at a Tecnai F20G2. TEM foils were prepared with a twin-jet electro polishing device, using an electrolyte composed of 10 vol.% perchloric acid and 90 vol.% acetic acid, applying a voltage of 58 V. APT specimens were electro-polished with the standard micro-polishing methods. APT analyses were performed using a Local Electrode Atom Probe (Cameca, LEAP$^{TM}$ 3000X HR) system in voltage pulsing mode at a specimen temperature of ~ 60 K and a pulse fraction of 15 %[17].

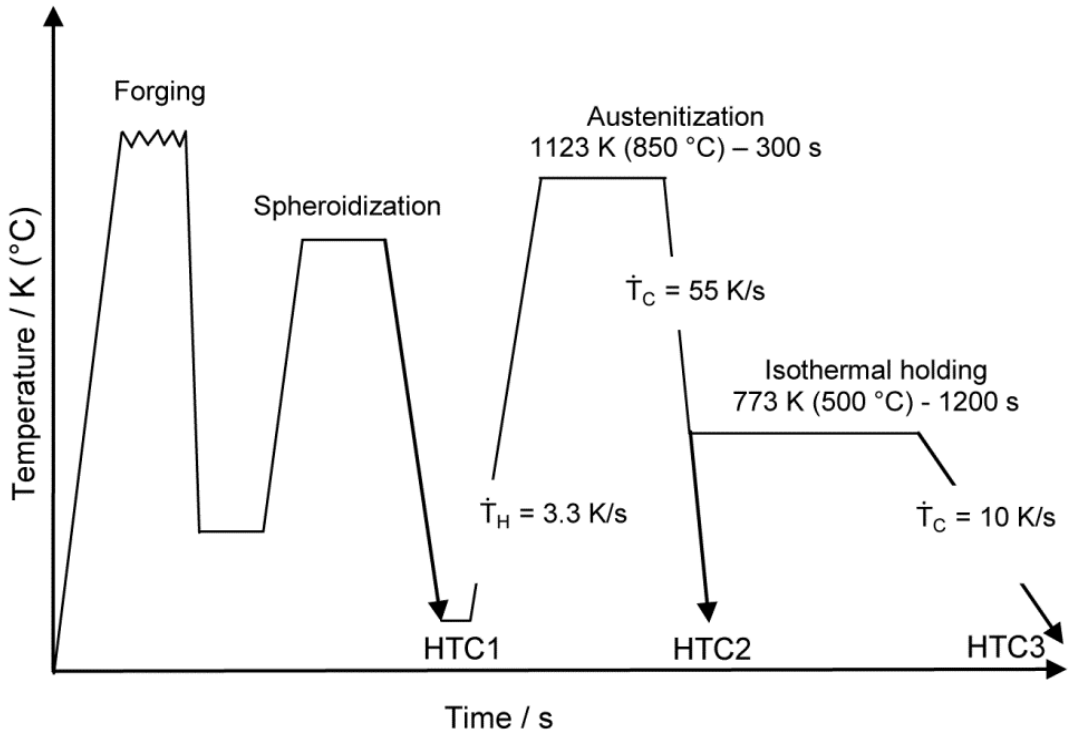

Fig. 1. Heat treatment cycle of the investigated steel 100Cr6

In order to interpret and validate the experimental data, thermodynamic calculations were performed using the software Thermo-Calc with the database TCFE7 and DICTRA with the mobility database MOBFE2[18]. The thermodynamic software Thermo-Calc is based on the CALPHAD method[19]. The DICTRA software is based on the assumption of diffusion controlled reactions and of local equilibrium (LE) at the moving interface, see e.g.[20-22]. The local equilibrium requires equal chemical potentials of all elements on either side of the interface. The numerical values of the chemical potentials are defined by the mass balance condition.

## 3. Results and discussions

### 3.1 Microstructure

#### 3.1.1 Morphology, phase fraction and size distribution of spheroidized carbides

The microstructures and corresponding spheroidized particle size distributions in 100Cr6 at different heat treated conditions are shown in Fig. 2. Fig. 2 (a) reveals the ferritic matrix with spheroidized carbides (particles of bright contrast in SEM image) after spheroidization at 1073 K (800 °C) for 7200 s (HTC1). The volume fraction of the spheroidized carbide particles was estimated from the area fraction in the SEM image and was obtained from measurements of 2000 particles for each heat treatment condition. By using metallographic methods (i.e. grinding, polishing, Klemm etching and following image analysis by contrast), the phase fraction of spheroidized carbides is determined to be 14 vol.%. The corresponding particle size distribution is displayed in Fig. 2 (b). Due to an incomplete dissolution into austenite, the spheroidized carbides are distributed inhomogeneously and some of the partially dissolved particles still exhibit elongated and irregular shapes. The spheroidized particles in the HTC1 condition have an average size of 0.63±0.02 µm. After austenitization at 1123 K (850 °C) for 300 s (HTC2), most of the spheroidized carbides are dissolved and only 3 vol.% remains. After quenching, the microstructure exhibits a martensitic matrix with partially dissolved spheroidized carbides (Fig. 2 (c)). The particles have become more spherical and smaller, with an average size 0.49±0.02 µm. After austenitization at 1123 K (850 °C) for 300 s followed by isothermal holding at 773 K (500 °C) for 1200 s (HTC3), the microstructure consists of a bainitic matrix with partially dissolved spheroidized carbides (see Fig. 2 (e)), where the corresponding particle size distribution is displayed in Fig. 2 (f). As is shown in Fig. 2 (e), elongated bainitic carbides within bainitic structure are formed during isothermal holding bainitic transformation in the HTC3 condition. Thus, the total fraction of carbides in bainitic matrix is more than that in martensitic matrix. However, the fraction of the spheroidized carbides in martensitic condition and bainitic condition is the same (Table 3). The martensitic matrix and bainitic matrix are determined by both heat treatment process and microstructure features. The martensitic matrix is gained by quenching in He after austenitization at a cooling rate of 55 K/s. The bainitic matrix is gained by isothermal holding at 773 K (500 °C) after austenitization. The quenched martensite show twinned plate features without carbides and the bainitic structure show bainitic ferrite lath with bainitic carbides which consists of inter-lath cementite substructure (Fig.3 (a)). The phase fractions and particle sizes of spheroidized carbides determined by metallographic analysis method (i.e. grinding, polishing, Klemm etching and following image analysis by contrast) are listed in Table 3. One can see that after 1123 K (850 °C) austenitization for 300 s followed by isothermal holding at 773 K (500 °C) for 1200 s (HTC3), the phase fraction of the spheroidized carbides remains constant as compared with that in HTC2 condition but the average size of the carbides becomes smaller.

Table 3. Phase fractions and particle sizes of partially dissolved spheroidized carbides under different heat treated conditions

| Heat Treated Condition (HTC) | HTC1 | HTC2 | HTC3 |
|---|---|---|---|
| **Phase fraction** | 14 vol.% | 3 vol.% | 3 vol.% |
| **Particle size (aver. diameter)** | 0.63±0.02 µm | 0.49±0.02 µm | 0.46±0.02 µm |

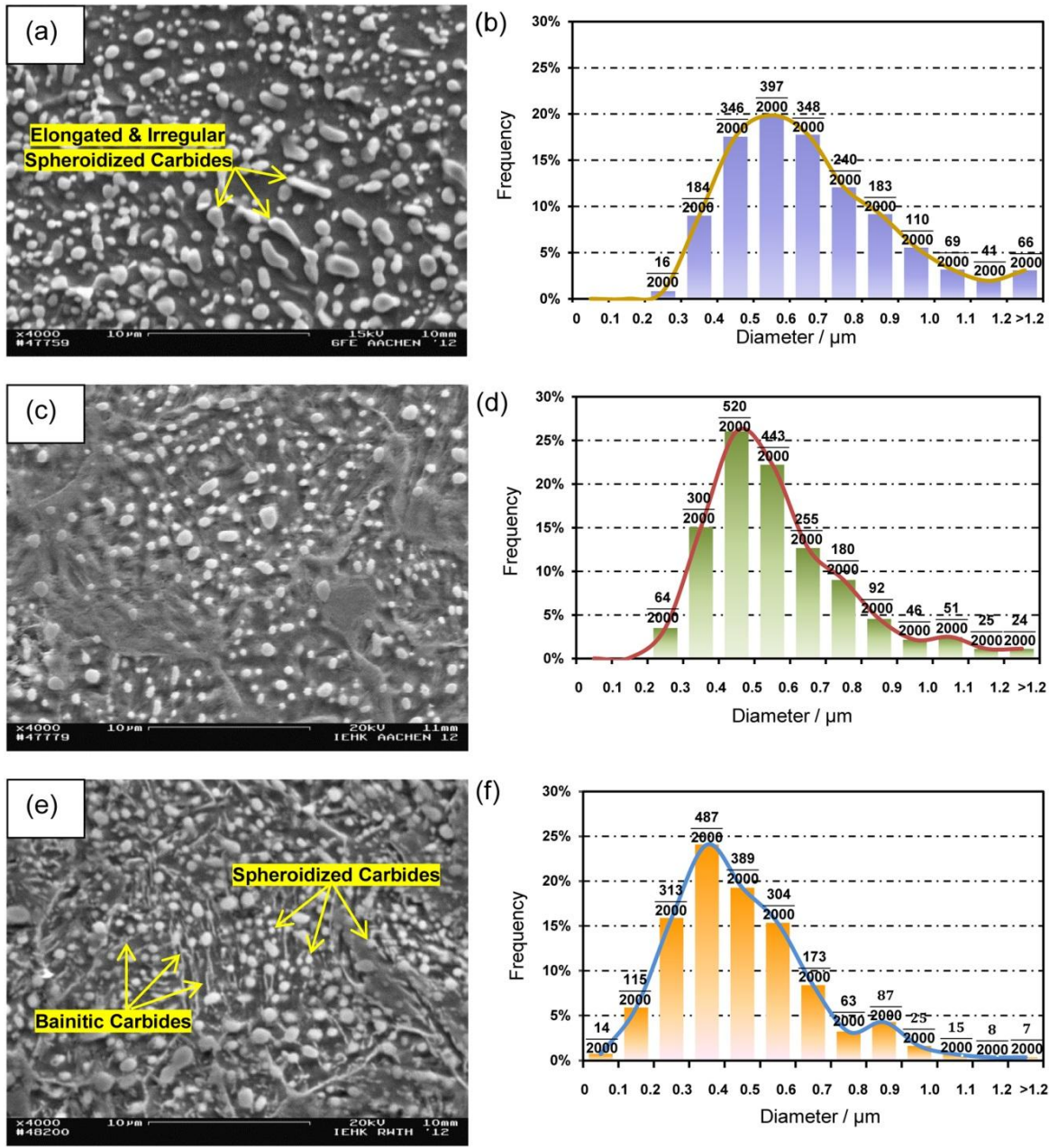

Fig. 2. shows the microstructure and spheroidized particle size distribution in 100Cr6 after different heat treatments: (a) ferritic matrix with spheroidized carbides under HTC1 conditions (b) size distribution of spheroidized carbides under HTC1 conditions (c) martensitic matrix with partially dissolved spheroidized carbides under HTC2 conditions (d) size distribution of partially dissolved spheroidized carbides under HTC2 conditions (e) bainitic matrix with partially dissolved spheroidized carbides under HTC3 conditions (f) size distribution of partially dissolved spheroidized carbides under HTC3 conditions (see Table 2 and Fig. 1).

### 3.1.2 Crystallography and EDX analysis on spheroidized carbides

Fig. 3 (a) displays the TEM micrograph of the bainitic microstructure in 100Cr6 obtained after isothermal heat treatment at 773 K (500 °C) for 1200 s (HTC3). During isothermal bainitic transformation at 773 K (500 °C), upper bainite microstructure is achieved. As is shown in the TEM bright field image in Fig. 3 (a), inter-lath cementite precipitates between bainitic ferrite laths within bainitic structure. As is mentioned above, 14 vol.% spheroidized carbides already exist before austenitization process (HTC1). During austenitization, the spheroidized carbides dissolve into austenite matrix. And after austenitization and quenching in He, 3 vol.% spheroidized carbides are detected in the steel (Fig. 2 (c), HTC2). During the following isothermal transformation at 500 °C (HTC3), no further dissolution of spheroidized carbides is expected and the fraction of spheroidized carbides under HTC3 condition maintains the same as that in HTC2 condition (Table 3). The partially dissolved spheroidized carbides remain in bainitic matrix. The TEM bright field image (Fig. 3 (a)) shows inter-lath $Fe_3C$ precipitates within bainitic structure and partially dissolved spheroidized carbides $(Fe,Cr)_3C$. The partially dissolved spheroidized carbides are proved to be $(Fe,Cr)_3C$ with the zone direction of [011] using the method of SAD (Selected Area Diffraction) in Fig. 3 (b). EDX chemical analysis further indicates the Cr content within $(Fe,Cr)_3C$ is about 12 wt.%, as shown in Fig. 4 point 4 measurement.

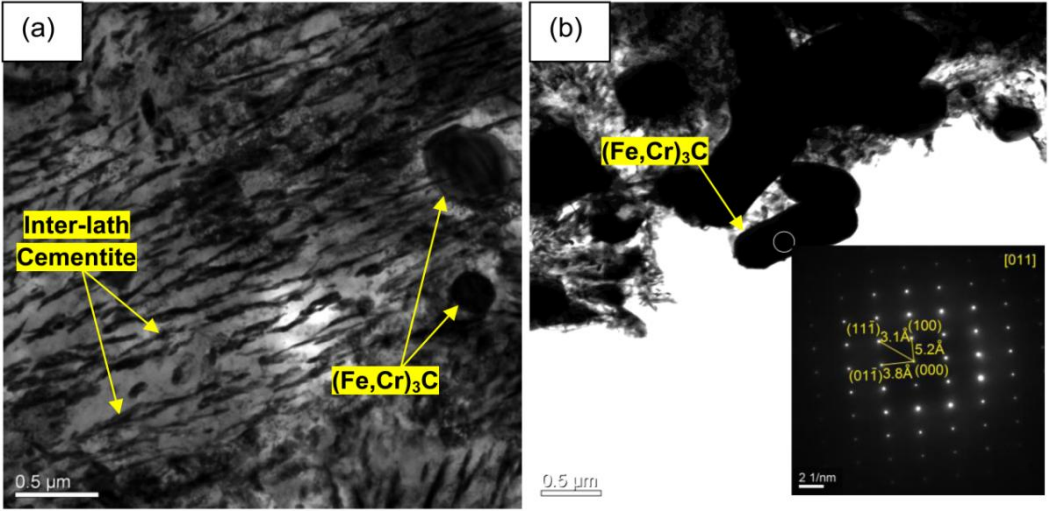

Fig. 3. (a)TEM bright field image showing the inter-lath $Fe_3C$ precipitation within bainitic structure and partially dissolved spheroidized carbide $(Fe,Cr)_3C$ in 100Cr6 isothermally heat treated at 773 K (500 °C) for 1200 s (HTC3, see Table 2 and Fig. 1); (b) Selected Area Diffraction (SAD) pattern showing partially dissolved spheroidized carbide $(Fe,Cr)_3C$ corresponding to [011] zone axis.

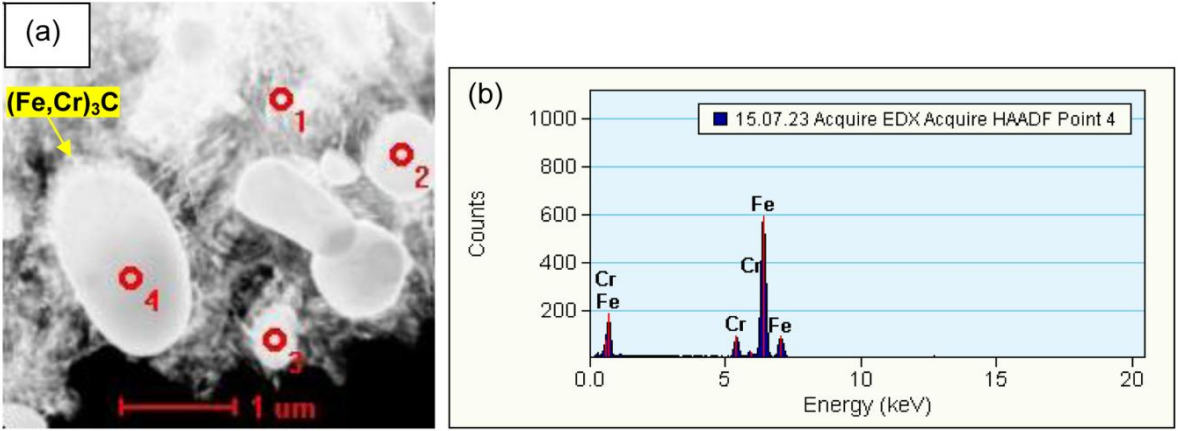

Fig. 4. (a) STEM image showing partially dissolved spheroidized carbide $(Fe,Cr)_3C$ in 100Cr6); (b) EDX chemical analysis at point 4 in Fig. 4(a).

## 3.2 Phase transformation kinetics

Fig. 5 (a) shows the dilatation curve of heating from room temperature at a rate of 3.3 K/s to 1123 K (850 °C) austenitization for 300 s followed by isothermal holding at 773 K (500 °C) for 1200 s (HTC3). Ferrite dissolution starting temperature $Ac_1$ is determined to be about 1033 K (760 °C) and its finishing temperature $Ac_3$ is at 1082 K (809 °C). Ferrite dissolution occurs quite fast which just experiences few seconds and this leads to the decrease of length. As shown in Fig. 5 (b), the length increases as a result of spheroidized carbide dissolution during austenitization at 1123 K (850 °C). The carbide dissolution rate is higher in the first few minutes and becomes slow gradually. In this process, both Cr and C diffusion play important roles. As is shown in the time-temperature profile of the phase transformations in 100Cr6 along the heat treatment cycle HTC3 in Fig.5 (c), ferrite dissolution is finished during heating process before austenitization at 1123 K (850 °C). Isothermal bainitic transformation at 773 K (500 °C) occurs quite fast at the first stage. Within the first 40 s, 90% bainite is formed. The incubation time is less than 1 s which means almost no incubation time is required for isothermal bainitic transformation in 100Cr6 at 773 K (500 °C).

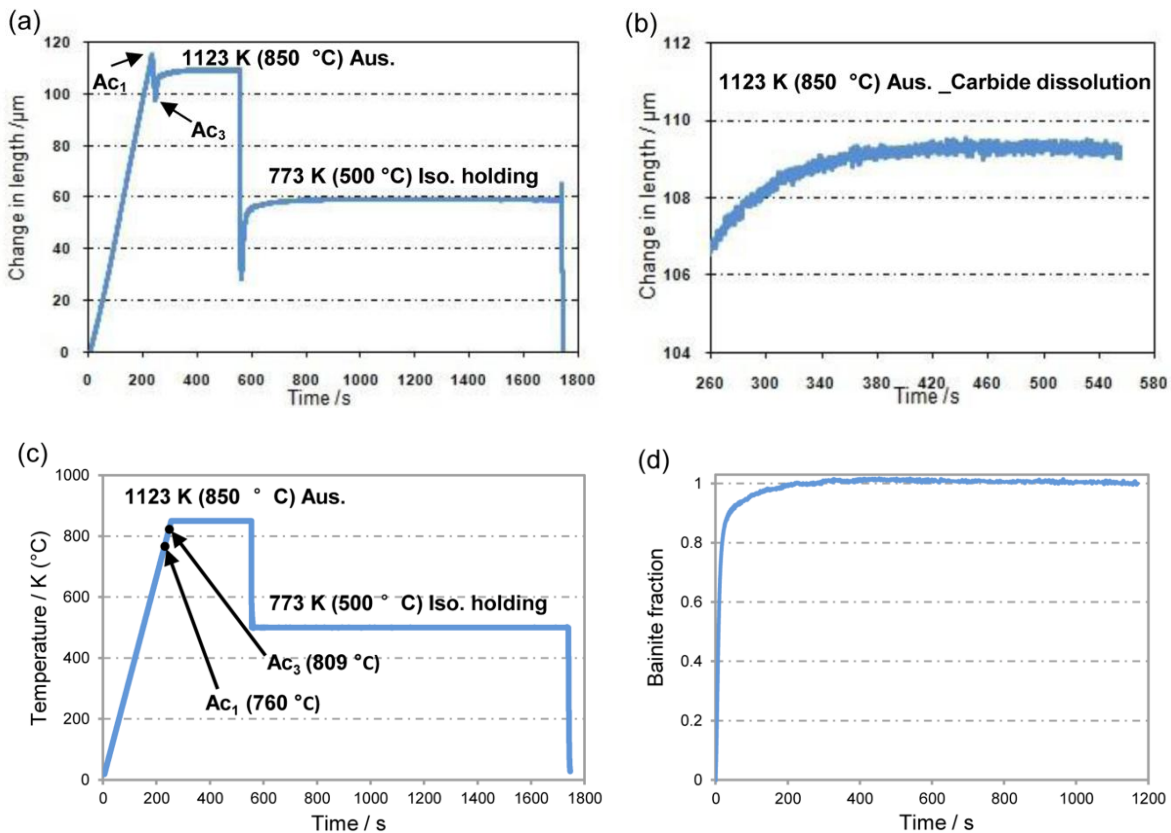

Fig. 5. (a) Change in length versus transformation time in 100Cr6 along the heat treatment cycle HTC3 (b) spheroidized carbide dissolution during austenitization at 1123 K (850 °C) (c) time-temperature profile of the phase transformations in 100Cr6 along the heat treatment cycle HTC3 (d) kinetics curve of isothermal bainitic transformation in 100Cr6 at 773 K (500 °C) for 1200 s (HTC3).

## 3.3 Atom Probe Tomography (APT)

3D atom maps obtained from the material isothermally heat treated at 773 K (500 °C) for 1200 s (HTC3) are shown in Fig. 6. The distribution of the carbon and alloy element atoms in the analysis volume is clearly non-uniform. Enrichment and depletion zones of C, Cr, Mn and Si can be recognized in the corresponding elemental maps.

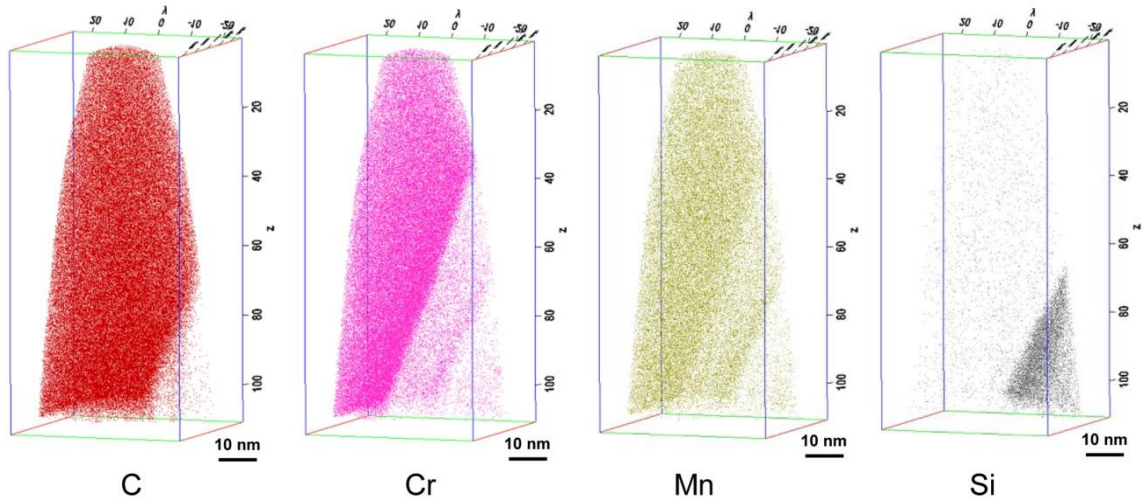

Fig. 6. 3-Dimensional atom maps obtained from the investigated material isothermally heat treated at 773 K (500 °C) for 1200 s (HTC3)

As is illustrated in Fig. 7, carbon atoms are distributed homogeneously in both partially dissolved spheroidized carbide (Fe,Cr)$_3$C and newly formed cementite. Carbon atoms exhibit a clear transition between cementite and bainitic ferrite. Due to the low solubility of carbon in ferrite, $α_B$ is depleted of carbon. In contrast, Si is mostly dissolved in the bainitic ferrite matrix and shows quite low solubility in cementite. Mn and Cr exhibit the same solute characteristics in partially dissolved spheroidized carbide (Fe,Cr)$_3$C, newly formed cementite at 773 K (500 °C) and bainitic ferrite matrix. Due to the fact that Cr and Mn have higher diffusivity and higher solubility in Fe$_3$C at high temperature, i.e. austenitization at 1123 K (850 °C), Cr and Mn atoms show a higher enrichment in spheroidized carbide (Fe,Cr)$_3$C than in newly formed cementite at 773 K (500 °C) and the smallest concentration in bainitic ferrite.

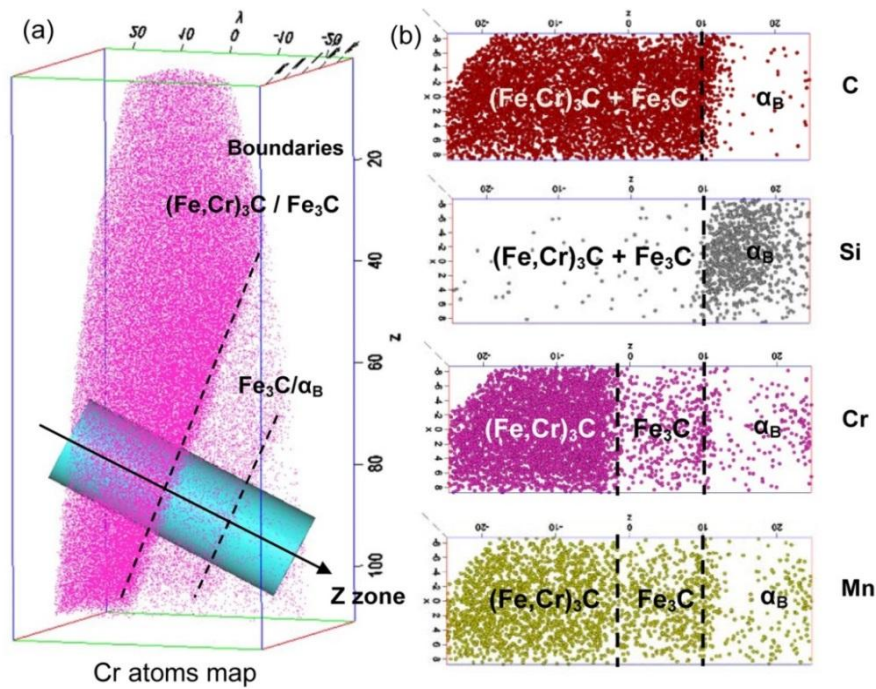

Fig. 7. (a) 3-Dimensional map of Cr atoms; (b) the selected box from Fig. 7 (a) showing the C, Si, Cr and Mn distributions; $(Fe,Cr)_3C$ is partially dissolved spheroidized carbide, $Fe_3C$ is the newly formed bainitic cementite.

Fig. 8 illustrates the 1D concentration profile (along the cylindrical region of interest in Fig. 7) showing the distribution of different elements, i.e. C, Si, Cr and Mn in partially dissolved spheroidized carbide $(Fe,Cr)_3C$ and in newly formed cementite at 773 K (500 °C) in contact with the bainitic ferrite matrix. The results in Fig. 8 (a) show a homogeneous distribution of the fast-diffusing element C, both in partially dissolved spheroidized carbides and in newly formed cementite during the isothermal holding process at 773 K (500 °C). Cr exhibits a gradual chemical gradient in the spheroidized carbides. The Cr concentration shows a maximum at the surface (16 at.%), gradually decreasing from the surface to the center in spheroidized carbides. Across the interface between partially dissolved spheroidized carbide $(Fe,Cr)_3C$ and newly formed bainitic $Fe_3C$, the Cr concentration shows a sharp decrease. This effect is attributed to the different solubility of Cr in austenite and ferrite as well as the different diffusivity of Cr at the austenitization temperature of 1123 K (850 °C) and the bainitic transformation temperature of 773 K (500 °C). At 773 K (500 °C), Cr has a very low diffusivity in bainitic ferrite which means that this element is essentially frozen. Table 4 shows the composition of cementite in equilibrium with ferrite and the diffusivities in cementite, ferrite and austenite. The data are derived from thermodynamic calculations using the software Thermo-Calc with the database TCFE7 and DICTRA with the mobility database MOBFE2.

Table 4a: Composition of cementite in equilibrium with ferrite (mole fractions)

| Temperature | x(Cr) | x(Mn) | x(Si) |
|---|---|---|---|
| **773 K (500 °C)** | 0.07813 | 0.01106 | $10^{-12}$ |
| **1123 K (850 °C)** | 0.06395 | 0.005455 | $4 \cdot 10^{-12}$ |

Table 4b: Diffusion coefficient ($m^2/s$)

| Phase | Temperature | Cr | Mn | Si |
|---|---|---|---|---|
| **Cementite** | 773 K (500 °C) | 8.268e-25 | 4.96e-25 | - |
| **Cementite** | 1123 K (850 °C) | 8.72e-19 | 5.23e-19 | - |
| **Ferrite** | 773 K (500 °C) | 5.23e-24 | 1.18e-22 | 1.37e-23 |
| **Ferrite** | 1123 K (850 °C) | 1.54e-15 | 3.26e-15 | 3.86e-15 |
| **Austenite** | 1123 K (850 °C) | 8.17e-18 | 1.035e-17 | 3.91e-17 |

Fig. 8 (b) represents the magnified curves of concentration profiles in a selected part of Fig. 8 (a). The partitioning characteristics of carbon and other alloying elements across the bainitic ferrite ($\alpha_B$) / cementite interface are revealed in Fig. 8 (b). The carbon partitions according to the local equilibrium between bainitic ferrite ($\alpha_B$) and cementite, while the substitutional elements Cr, Mn and Si redistribute over a short range and exhibit enrichment at the $\alpha_B/\theta$ interface. Si is mostly dissolved in the bainitic ferrite matrix and exhibits a relatively low solubility in cementite. Fig. 8 (b) shows that Si is highly enriched at the interface between the bainite and the cementite, which kinetically impedes the further growth of cementite. The data also reveal that there is no long range redistribution of the substitutional elements Cr, Mn and Si. The various interstitial and substitutional elemental partitioning features in the current APT investigation infer that cementite precipitate in upper bainite under non-partitioning local equilibrium (NPLE) mode. Moreover, cementite precipitates under a different mode during lower bainite formation. Our previous Atom Probe Tomography (APT)[23] reveals that cementite precipitates under negligible partitioning local equilibrium (NPLE) mode in upper bainite at 500 °C and ε carbide and cementite precipitate under paraequilibrium mode at 260 °C during lower bainite formation. Barrow and Rivera [24] investigated the cementite formation during bainite reaction at a low transformation temperature (about 230 °C) by means of energy-dispersive spectroscopy and reported that cementite formation during bainite reaction takes place under paraequilibrium.

Fig. 8 indicates that the partially dissolved spheroidized carbides during 1123 K (850 °C) austenitization may serve as nucleation sites for low temperature cementite and grow by the formation of a low Cr content cementite layer during 773 K (500 °C) bainitic isothermal holding process, which shows a relatively softer surface and harder inner part in the spheroidized particles. This sophisticated microstructure may lead to a better wear resistance and fatigue properties.

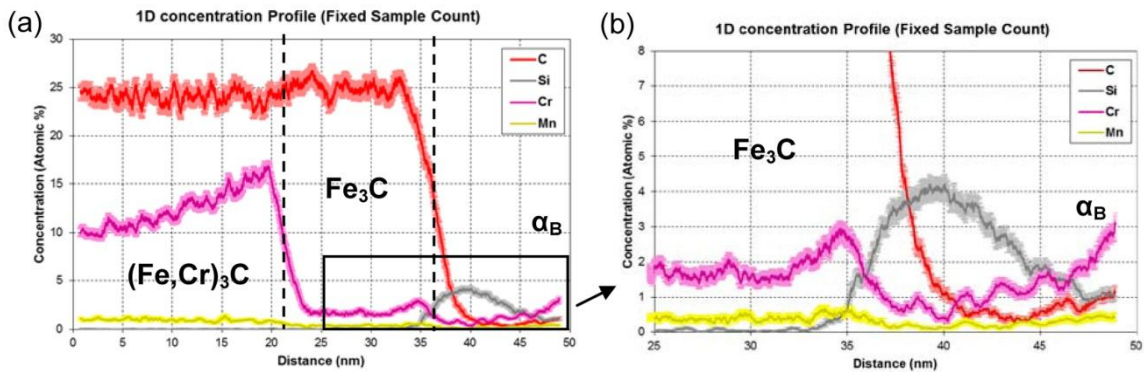

Fig. 8. 1-Dimensional concentration profile (Fixed Sample Count) showing the distribution of C, Si, Cr, Mn in partially dissolved spheroidized carbide (Fe,Cr)$_3$C, newly formed cementite at 773 K (500 °C) and bainitic ferrite matrix, which is obtained from the investigated material isothermally heat treated at 773 K (500 °C) for 1200 s (HTC3); Fig. 8 (b) represents the magnified curves of concentration profiles in the selected area in Fig. 8 (a)

### 3.4 DICTRA Simulations

In the case of systems with fast diffusing interstitial elements like C, the local equilibrium may define two different growth regimes: a fast reaction with no partitioning of substitutional components (NPLE) or a slow reaction with partitioning of all elements (LE). In the case of one substitutional sublattice (e.g. in ferrite, austenite, cementite) the fraction of component i on the sublattice is denominated u-fraction. In terms of the number $N_i$ of atoms of type i, or in terms of the mole fraction $x_i$, the u-fractions are given by $u_i = \frac{N_i}{\sum_{j \neq C} N_j} = \frac{x_i}{1-x_C}$. The NPLE tie-line is defined by the new phase exhibiting the same u-fraction of substitutional elements as the matrix. There is a strict thermodynamic condition for the NPLE reaction to be possible. If the amount of interstitial element is lower in the new phase than in the matrix (e.g. formation of ferrite from austenite) the activity of the interstitial component in the new phase must be higher than in the matrix, so that the interstitial component is transported into the matrix phase. In the opposite case, e.g. the formation of cementite from ferrite, the C-activity within cementite with the same u-fraction of substitutional elements as the ferritic matrix must be lower than in the matrix, leading to a C-flux towards cementite.

In order to keep the diffusion simulations reasonable in time, only the major components C, Cr, Mn, Si were considered. The calculations were thus performed for the alloy composition Fe-1C-1.4Cr-0.23Mn-0.3Si (wt.%), i.e. Fe-4.468C-1.445Cr-0.2247Mn-0.5733Si (at.%). The substitutional sublattice composition of ferrite is thus $u(Cr) = 0.015126$, $u(Mn) = 0.002352$, $u(Si) = 0.006001$. Let us denote by $M_3C^*$ the cementite with the same u-fraction of Cr and Mn as the alloy. According to the database TCFE7 cementite dissolves only a negligible amount of Si. The composition of cementite is then $X_{Cr} = 0.011345$, $X_{Mn} = 0.001764$ $X_C = 0.25$. In Table 5, the C-activities of the ferritic matrix and of $M_3C^*$ derived from Dictra calculations are given for the temperature interval from 773 K (500 °C) to 1073 K (800 °C). In all cases, the C-activity at the interface $M_3C^*$ / ferrite is lower than in the ferritic matrix leading to C-diffusion from the matrix towards the carbide. Therefore, the formation of cementite is expected to occur according to the NPLE condition in the whole temperature interval.

Table 5. C-activities with reference to graphite at the given temperatures

| Temperature | $a_C$ (ferrite) | $a_C$ ($M_3C^*$) |
|---|---|---|
| **1073 K (800 °C)** | 2.5247E+01 | 9.3864E-01 |
| **973 K (700 °C)** | 8.0311E+01 | 1.2067E+00 |
| **873 K (600 °C)** | 3.3301E+02 | 1.7835E+00 |
| **773 K (500 °C)** | 1.9950E+03 | 3.0876E+00 |

The DICTRA simulations were performed assuming a spherical cell with 0.6 µm radius. This corresponds to the experimental observation that the average particle diameter is 0.63 µm representing 14 vol.% of the total volume. The simulation was started with a cementite nucleus of radius r = 1 nm. The nucleus composition can be obtained from Thermo-calc, see Table 6. This composition can be entered if the particle is defined as present. If the particle is not yet present and forms only after reaching a certain driving force for precipitation, then DICTRA derives this composition automatically. The volume fraction of cementite obtained from DICTRA simulations at 773 K (500 °C), 973 K (700 °C) and 1073 K (800 °C) are given in Fig. 9. The most important result of these simulations is the extremely short time of less than one second for the precipitation of cementite. After 0.1 s cementite has already reached the plateau of its final volume fraction.

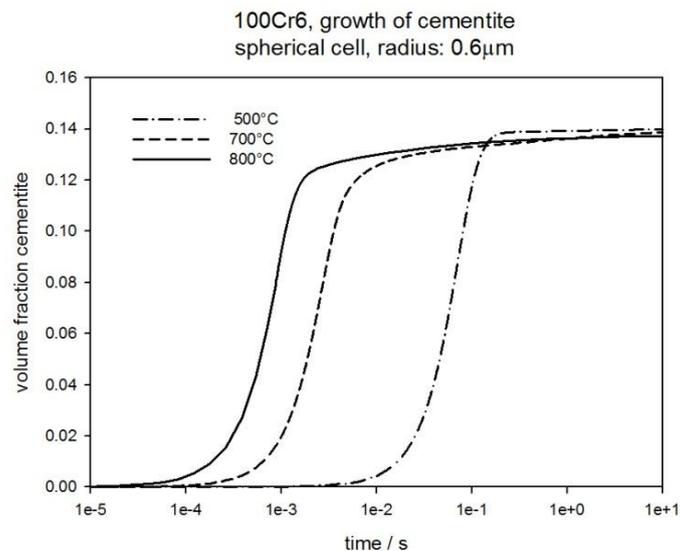

Fig. 9. Growth of cementite in ferrite according to NPLE

Table 6. Composition of the critical cementite nucleus

| Temperature | X(Cr) | X(Mn) | X(Si) |
|---|---|---|---|
| **1073 K (800 °C)** | 1.10813E-01 | 7.02977E-03 | 7.50000E-13 |
| **973 K (700 °C)** | 1.42356E-01 | 1.16464E-02 | 7.50000E-13 |
| **873 K (600 °C)** | 1.92662E-01 | 1.95503E-02 | 7.50000E-13 |
| **773 K (500 °C)** | 2.74839E-01 | 3.16540E-02 | 7.50000E-13 |

The composition profiles in Fig. 10 show that the growth of cementite does indeed proceed according to the NPLE reaction with respect to Cr and Mn. Si cannot be dissolved in cementite. Therefore a spike of Si is pushed ahead of the moving interface. The height of the spike varies with time and corresponds to LE. Cementite has reached its volume fraction of about 14 vol.% in about 0.1 s. Similar profiles are obtained at 973 K (700 °C) and 1073 K (800 °C). The Cr- and Mn-content of the growing cementite corresponds to the experimental results obtained with Atom Probe Tomography of the chemical concentration in cementite precipitates.

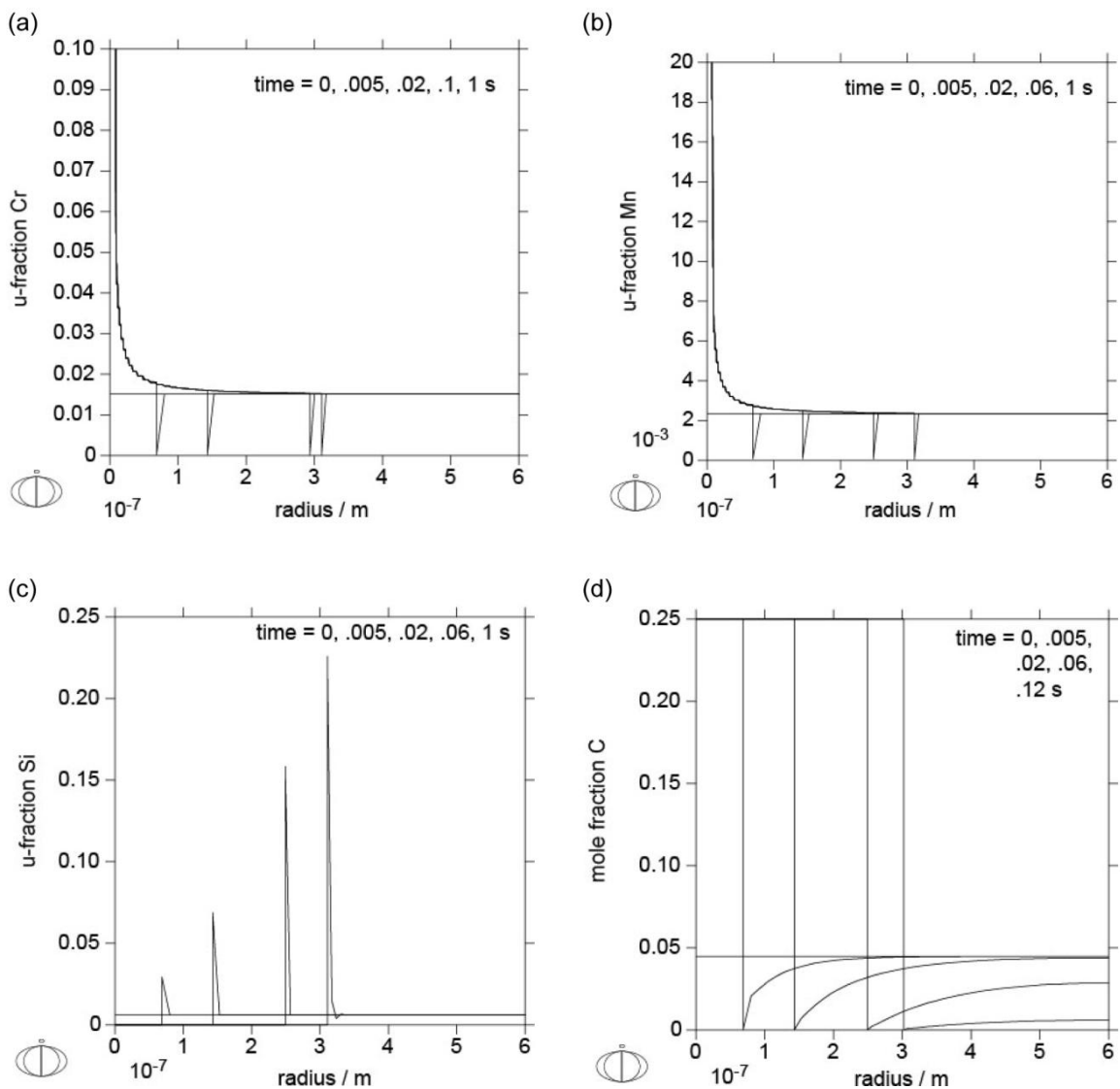

Fig.10. Composition profiles during growth of cementite in ferrite at 773 K (500 °C) for various time steps.

The results of annealing at 1073 K (800 °C) for 2 h and 5 h are shown in Fig. 11. The experimental results of the austenitization at 1123 K (850 °C) show that, in order to get good agreement, the mobility of the substitutional elements in cementite has to be increased by a factor of about five compared to the data in MOBFE2. Therefore, in Fig. 11 the results for both sets of mobilities are shown. The comparison between Fig. 11a and 11b already shows that the difference in mobilities creates significantly different composition profiles. The starting configuration of these simulations was cementite formed by the NPLE reaction. This corresponds to the experimental situation with cementite formed during heating or cooling. Due to the NPLE reaction cementite has grown with a much lower Cr- and Mn-content compared to global equilibrium. Consequently, a reaction towards equilibrium at 1073 K (800 °C) requires an increase of these elements in cementite. This leads to diffusion of Cr and Mn from ferrite into cementite. Si diffuses into the opposite direction and dissolves in ferrite. The Si-spike is completely removed already after about 30 s, see Fig. 12b. This is due to the much higher mobility of Si in ferrite compared to the mobility of substitutional elements in cementite. In conclusion, the treatment at 1073 K (800 °C) leads to a redistribution of substitutional elements, but there is no further growth within the 2 h heat treatment.

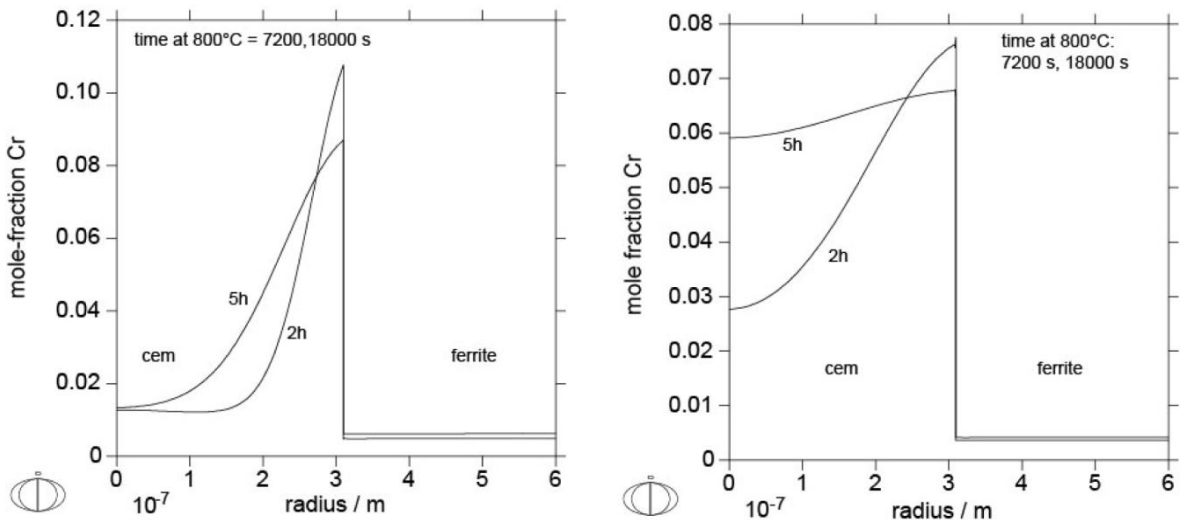

(a) Mobilities of Cr according to MOBFE2    (b) Mobilities of Cr increased by a factor of 5
Fig. 11. Cr-composition profiles formed during annealing at 1073 K (800 °C)

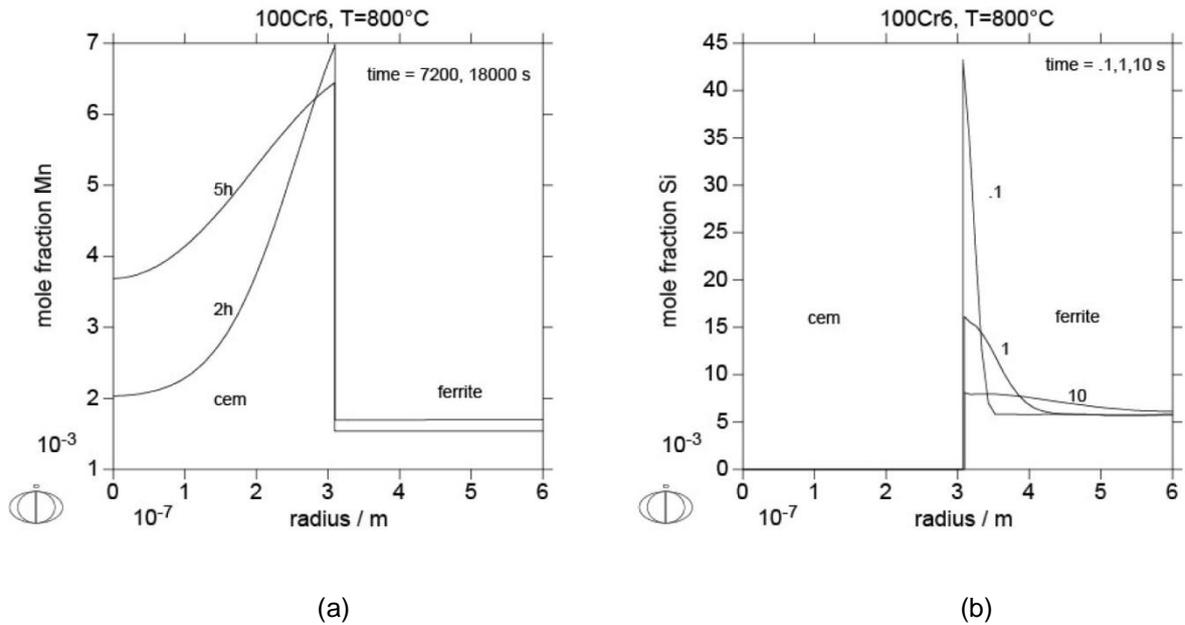

Fig. 12. Mn, Si-composition profiles formed during annealing at 1073 K (800 °C)
( Mobilities of Cr, Mn, Si increased by a factor of 5)

At 1123 K (850 °C) ferrite transforms into austenite. The starting C-content in austenite is very low since it is given by that of ferrite at 1073 K (800 °C). Consequently, cementite dissolves and austenite absorbs the C. In order to simulate this reaction, the starting composition profiles were taken from the result at 1073 K (800 °C). The composition profiles of the ferrite region at 1073 K (800 °C) define the composition in the austenite region at 1123 K (850 °C). The results of the simulation of a treatment at 1123 K (850 °C) for 300 s are shown in Fig. 13 for the two cases, mobilities according to MOBFE2 (Fig. 13a) and increased by a factor of 5 (Fig. 13b). Fig 14 shows a blow-up of the Cr content near the interface with a scaling such that a comparison can be made with the experimental data in Fig. 8. The result obtained with the mobility of the database MOBFE2 shows a higher level of Cr at the interface, a steeper decrease of the Cr-profile within cementite, and a radius of r = 0.268 μm which is slightly higher than the experimental average particle size. The results obtained with a change of mobility by a factor of 5 show good agreement with the experimental compositional data measured by Atom Probe Tomography (APT), and the particle radius after shrinking is r = 0.247 μm, in excellent agreement with the experimentally observed average particle diameter of d = 0.49 μm. The shrinkage of cementite sets Mn, Cr and C free for dissolution in austenite. Both APT observations (see Fig. 8) and DICTRA calculations (see Fig. 13) show a gradual chemical gradient of Cr inside the spheroidized carbides. Cr exhibits a maximum concentration at the surface of spheroidized carbides and decreases gradually from surface towards the core. The calculated results indicate the minimum concentration of Cr at the core of spheroidized carbides down to a level of about 2 at.%.

The subsequent treatment at 773 K (500 °C) leads to the growth of cementite according to the NPLE regime. The Cr concentration in ferrite corresponds to that of austenite at 1123 K (850 °C). Close to the interface ferr / cem the Cr concentration is about $x_{Cr}^{\alpha}$ = 0.024, and the C-concentration is $x_{C}^{\alpha}$ = 0.029. Cementite will thus grow with a Cr u-fraction of $u_{Cr}^{cem}$ =

0.0247. The mole fraction of Cr measured in cementite is then $x_{Cr}^{cem} = x_{Cr}^{cem} \cdot 0.75 = 0.0185$. This is in accordance with the result in Fig. 8.

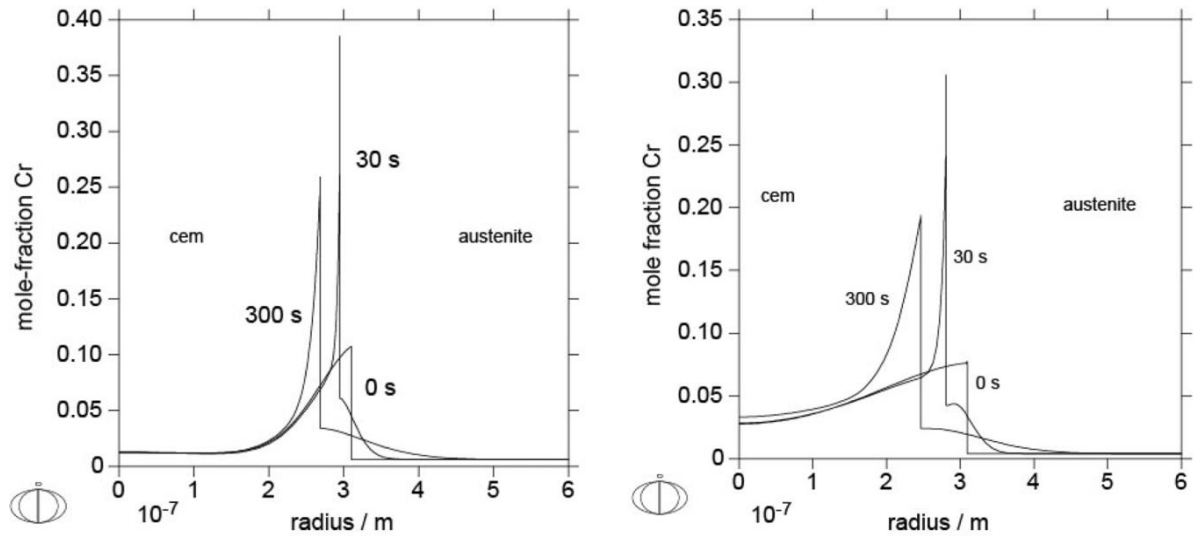

(a) Mobilities of Cr, Mn, Si according to MOBFE2   (b) Mobilities of Cr, Mn, Si increased by a factor of 5

Fig. 13. Composition profiles during austenitization at 1123 K (850 °C)

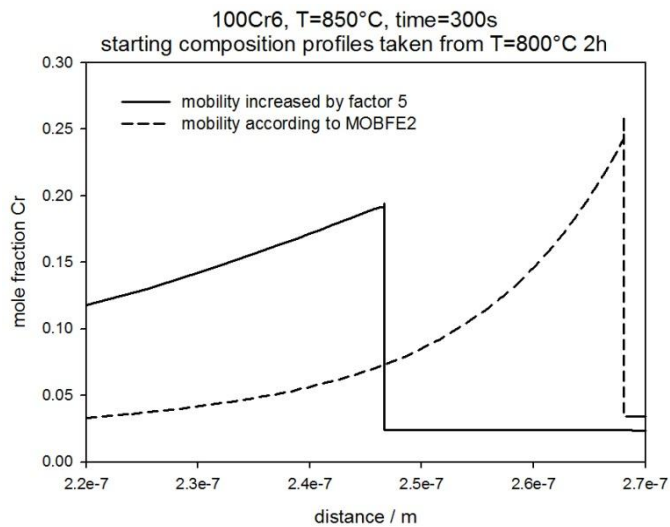

Fig. 14. Blow-up of the Cr concentration profiles in Fig. 13, temperature 1123 K (850 °C), time 300 s. The scaling of the x-axis has been selected similar to that of the experimental results in Fig. 8(a).

## 4. Conclusions

In our present work, spheroidized carbides in a high carbon bearing steel 100Cr6 were characterized with respect to their morphology, phase fraction, size distribution, crystallography and kinetics of carbide dissolution behavior, applying SEM, TEM and APT. The spheroidized carbide dissolution kinetics was analyzed using dilatometry. The chemical gradient of C, Cr, Mn and Si within spheroidized carbides, cementite, bainitic ferrite matrix and the atom partitioning behaviors across phase boundaries were investigated at the atomic scale. By using the APT technique, we gained deeper insights into the natures of spheroidized carbides and transformation dynamics of carbon, chromium and other alloy element atoms. The results can be summarized as follows:

1. 14 vol.% spheroidized carbides (Fe,Cr)$_3$C are formed during soft annealing and 3 vol.% remain after dissolution into the austenite matrix during austenitization at 1123 K (850 °C) for 300 s.
2. The fast diffusing element C is distributed homogeneously in both, the partially dissolved spheroidized carbides and in the cementite formed during the 773 K (500 °C) bainite stage.
3. Cr exhibits a gradual chemical gradient inside the spheroidized carbides. The Cr concentration exhibits a maximum value (16 at.%) at the surface of the spheroidized carbides, gradually decreasing from surface to the core down to a level of about 2 at.%.
4. Spheroidized carbides remaining partially dissolved after austenitization at 1123 K (850 °C) may exist as nucleation site for lower temperature cementite within bainite. The latter grows with a relatively lower Cr concentration (2 at.%) during 773 K (500 °C) bainitizing.
5. Si exhibits a large enrichment at the growth front of cementite, which hinders the coarsening of cementite particles.
6. DICTRA calculations reveal that cementite precipitation at 773 K (500 °C) occurs under non-partitioning local equilibrium condition (NPLE) mode and show good agreement with the experimental compositional distributions detected by Atom Probe Tomography (APT).

**Acknowledgements**

This work has been performed within the Interdisciplinary Centre for Advanced Materials Simulation (ICAMS) at Ruhr University Bochum. ICAMS gratefully acknowledges funding from ThyssenKrupp Steel Europe AG, Bayer Material Science AG, Salzgitter Mannesmann Forschung GmbH, Robert Bosch GmbH, Benteler Steel/tube Management GmbH, Bayer Technology Services GmbH and the state of North Rhine-Westphalia as well as the European Commission in the framework of the European Regional Development Fund (ERDF). Authors would like to express their many thanks to Dr. Christoph Somsen in Ruhr-University Bochum for his support of TEM experiment.